\documentstyle [prl,multicol,aps,psfig]{revtex}
\begin{document}
\title{Quantum limited sensitivity of SET-based displacement detectors}
\author{D. Mozyrsky, I. Martin, and M. B. Hastings}
\address{Theoretical Division, Los Alamos National Laboratory, Los Alamos, NM 87545, USA}
\date{Printed \today}
\maketitle
\begin{abstract}
We consider a model of a quantum-mechanical resonator capacitively coupled to a
single electron transistor (SET). The tunnel current in the SET is modulated by
the vibrations of the resonator, and thus the system operates as a displacement
detector. We analyze the effect of the back-action noise of charge fluctuations
in the SET onto the dynamics of the resonator and evaluate the displacement
sensitivity of the system. The relation between the ``classical" and ``quantum"
parts of the SET charge noise and their effect on the measured system are also
discussed.
\end{abstract}

\pacs{PACS Numbers: XXXXX}
\vspace{-9 mm}
\begin{multicols}{2}

Micro-mechanical resonators have been used as ultra-sensitive force detectors
in a number of experimental applications, ranging from Atomic Force Microscopy
to Magnetic Resonance Force Microscopy\cite{MRFM} to experiments on Casimir
force detection\cite{Casimir}.  Recently, mechanical resonators with
vibrational eigenfrequencies ($\nu$) of the order of 1 GHz have been
fabricated\cite{Roukes1G}.  At low temperatures ($h\nu
> k_BT\sim$ 50 mK),  these resonators provide an example of a man-made system
that can be used to test the basic principles of quantum mechanics at the
macroscopic level.

The standard cantilever displacement measurement schemes are based on laser
interferometry, and can reach the levels of sensitivity of the order
$10^{-4}$\AA$/\sqrt{{\rm Hz}}$. This level of sensitivity requires, however,
high laser power that may not be compatible with the ultra-low temperature
operation.  This limitation provided the motivation to explore alternative
$electical$ measurement schemes\cite{BW,expt,MM}.  In particular, Blencowe and Wyborne\cite{BW}
have suggested based on a semiclassical analysis that by capacitively coupling
the cantilever to a Single Electron Transistor (SET), it is possible to achieve
the sensitivity better than the zero-point-motion uncertainty.  More recently,
two of us\cite{MM} have found based on a fully quantum mechanical description
of the quantum measurement of a cantilever using a quantum point contact (QPC),
that the apparatus back action (current shot noise that induces force noise on
the cantilever) fundamentally limits the displacement sensitivity and leads to
a quantum-to-classical transition in the oscillator dynamics.  Due to the
resonant nature of transport through SET, it is expected to have significantly
higher displacement sensitivity that QPC, and hence is more attractive from the
experimental standpoint.  Here, we analyze the fundamental sensitivity limits
of an SET-based detection scheme.  We find that the higher ``classical''
sensitivity comes at the expense of drastically increased back-action (island
charge noise), which leads to large rms fluctuations of the cantilever and to
deterioration of the oscillator quality factor in the experimentally most
attractive $threshold$ regime.  The optimal sensitivity is achieved in the {\em
co-tunneling} regime in which the SET scheme becomes equivalent to the
QPC-based detection.  Similar conclusions have also been reached in the study
of SET charge sensitivity ~\cite{Averin}.  We find that under no circumstances
is it possible to exceed the standard quantum limit using continuous SET-based
detection.

\begin{figure}
{\centering{\psfig{figure=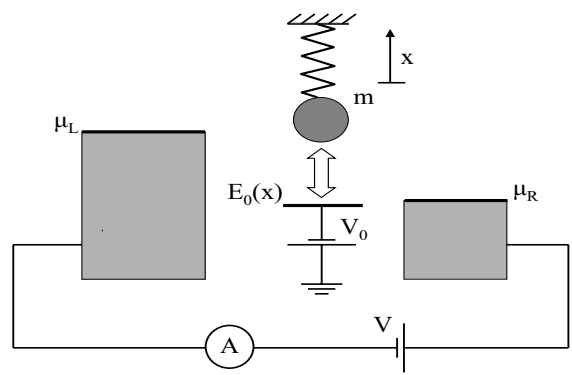,height=2. in,width=3.1
in,angle=90}}}\caption{Schematics for the model setup.}
\end{figure}

The model is schematically presented in Fig. 1. The measuring apparatus is a
single electron transistor: a quantum dot coupled to the leads via tunnel
junctions. For simplicity we assume that the dot contains a single resonant
level. The mechanical system, in the following referred to as an oscillator,
which can be either a micro-mechanical resonator or a localized phonon mode, is
capacitively coupled to the resonant level. Therefore, the displacement of the
oscillator can alter the position of the resonant level with respect to the
chemical potentials in the leads and thus can affect the tunnel current through
the device.

The Hamiltonian of the model can be written as
\begin{equation}
H = H_{\rm leads} + H_{\rm osc} + H^{\prime}\, ,\label{a0}
\end{equation}
where the first two terms are the Hamiltonians of the electrons in the leads
and the oscillator respectively, $H_{\rm leads} = \sum_{q, n=L,R} \epsilon_{qn}
c^{\dag}_{qn}c_{qn}$ and $H_{\rm osc} = (m/2)\partial_x^2 + V(x)$. Here
$c^{\dag}_{q,n=L(R)}\ (c_{q,n=L(R)})$ creates (annihilates) an electron with a
quantum number $q$ in the left (right) lead, $m$ and $\omega_0$ are the
oscillator effective mass and frequency, and $x$ is the oscillator coordinate.
Here and in the following we set both Planck's constant $\hbar$ and the
electron charge $e$ equal to unity, unless stated otherwise. The single
particle states $E_{qn}$ are filled up to chemical potentials in the leads,
$\mu_L$ and
$\mu_R$, which are biased by external voltage, $\mu_L - \mu_R = V$, see Fig. 1. For simplicity we
assume zero external temperature. The Hamiltonian $H^{\prime}$ includes both
electron tunneling and modulation of the position of the resonant level by the
oscillator:
\begin{equation}
H^{\prime} = \sum_{q, n=L,R} T_n(d^{\dag} c_{qn} + c^{\dag}_{qn}d) + E_0(x){\hat n}_0\, .\label{a1}
\end{equation}
In the first, tunneling, term of the Hamiltonian~(\ref{a1}), the operator
$d^{\dag}_0\ (d)$ creates (annihilates) an electron in the resonant level, ${\hat
n}_0 = d^{\dag} d$, and the tunneling amplitude $T_n$ is assumed to be
independent of the single particle states in the leads. We assume that the
energy of the resonant level $E_0(x)$  depends linearly on the oscillator's
coordinate $x$, i.e. $E_0(x)= \epsilon_0+\lambda x$. The unperturbed position
of the resonant level
$\epsilon_0$ can be varied by appropriately adjusting the gate voltage $V_0$, see Fig. 1.  The
parameter $\lambda$ physically represents an effective electric field in the
capacitor formed by the oscillator and the quantum dot.

We use the Keldysh-Feynman-Vernon formalism~\cite{Keldysh,Feynman} to determine
the evolution of the oscillator under the influence of the tunneling electrons.
We define a scattering operator for the oscillator alone, i.e. with electronic
degrees of freedom traced out:
\begin{equation}
{\cal S}_{\rm osc} = {\rm Tr}_{\rm el}\left[\rho_{el} {\cal T}_{\rm c}\,{\cal S}(-\infty,\infty)
{\cal S}(\infty,-\infty)\right]/{\rm Tr}\left[\rho_{\rm el}\right]\, .\label{a2}
\end{equation}
In Eq.~(\ref{a2}) ${\cal S}(\infty,-\infty)$ and ${\cal S}(-\infty,\infty)$ are scattering
operators for the full system, ${\cal S}(\infty,-\infty) = \exp[{-i\int_{-\infty}^\infty}H dt]$,
where $H$ is defined in Eqs.~(\ref{a0},\ref{a1}), and the operator ${\cal T}_{\rm c}$ denotes time
ordering along the Keldysh contour. The density matrix of the unperturbed electrons is the direct
product of the uncoupled density matrices of electron reservoirs in the leads ($\rho_L$ and $\rho_R$)
with an empty electron state in the resonant level ($\rho_D=d d^{\dagger}$), $\rho_{\rm el}=
\rho_L\otimes\rho_R\otimes\rho_D$. Eq.~(\ref{a2}) implies that at $t=-\infty$ the leads, the
resonant level, and the oscillator are uncoupled and that the interaction,
$H^{\prime}$, is
switched on adiabatically at $t>-\infty$.

In what follows we assume that the coupling constant $\lambda$ is small, while the tunnelling
amplitudes $T_L$ and $T_R$ need not be small. Then, ${\cal S}_{\rm osc}$ can be written explicitly
as a functional integral over the oscillator coordinate as
\begin{eqnarray}
{\cal S}_{\rm osc} &=& \int {\cal D}\,x_{\rm c}\exp{\left[i\int_{\rm c} dt
{\cal L}^{\prime}_{\rm osc}\right]}
~~~~~~~~~\nonumber\\
&\times&\exp{\left[-{\lambda^2 \over 2} \int_{\rm c} dt_1dt_2 x(t_1)x(t_2)
K(t_1 - t_2) + ...\right]}\, .\label{a3}
\end{eqnarray}
In this work we limit ourselves to $O(\lambda^2)$ contribution to the effective action of the
oscillator. The higher orders in the expansion (denoted by $...$ in Eq.~(\ref{a3})) are unimportant
in the limit of strong tunneling as will be seen below.

The first order contribution in $\lambda$, i.e. interaction of the oscillator with average charge
$\langle{\hat n}_0\rangle$ in the dot, is included in the Lagrangian of the oscillator in
Eq.~(\ref{a3}), ${\cal L}^{\prime}_{\rm osc} = {\cal L}^{\rm (bare)}_{\rm osc} -
\lambda\langle{\hat n}_0\rangle_{\rm el} x$. The charge $\langle{\hat n}_0\rangle$ is related to
the Fourier transform of the {\it renormalized} single particle Green's function $G_D(t^\prime-t) =
-i\langle {\cal T}_{\rm c}\, d(t)d^\dag(t^\prime)\rangle_{\rm el}$ as
$\langle n_0 \rangle_{\rm el} = (1 / 2\pi i)\int d\omega
\,G_D^{-+}(\omega)$.  The averaging denoted by $\langle\ \rangle_{\rm el}$, is taken with respect to
the exact stationary state of the electronic subsystem alone, i.e. decoupled
from the oscillator. The renormalization of $G_D(t)$ by the tunneling
transitions can be obtained by a standard calculation~ \cite{Juaho}. Following
the notation of Ref.~\cite{Keldysh,Rammer} we define a matrix Green's function
$G_D^{ij}(t_2-t_1) = -i \langle d(t_1^j) d^\dag(t_2^i)\rangle_{\rm el}$, where
$t_1^i$ and $t_2^j$ can either be on the same or different Keldysh contours,
i.e.
$i,j=\pm$. We also introduce the unperturbed Green's functions of the electrons in the left and the
right leads $G_{qn,0}^{ij}(t_1 -t_2)= -i\langle {\cal T}_{\rm c}\,
c(t_1^j)_{qn}c^\dag(t_2^i)_{qn}\rangle_{\rm el,0}$, $n=L,R$, and the unperturbed Green's function
of the dot electron $G_{D,0}^{ij}(t_1 -t_2)= -i\langle {\cal T}_{\rm c}\,
d(t_1^j)d^\dag(t_2^i)\rangle_{\rm el,0}$. The time ordered and anti-time ordered Green's functions,
i.e. with time arguments on forward and return branches respectively can be expressed in terms of
the Green's functions with time arguments on different branches as $G^{++}(t) = \Theta(t)G^{-+}(t)
+ \Theta(-t)G^{+-}(t)$ and $G^{--}(t) = \Theta(t)G^{-+}(t) + \Theta(-t)G^{+-}(t)$, where
$\Theta(t)$ is a unit step function~\cite{Keldysh,Rammer}. Then, by solving the Dyson equation
$G_D^{ij} = G_{D,0}^{ij}+G_{D,0}^{ik} \Sigma^{kl} G_D^{lj}$, where self-energy $\Sigma^{ij} =
\sum_{q, n=L,R} T_n^2 G_{qn,0}^{ij}$, after straightforward calculation we obtain:
\begin{mathletters}
\label{a4}
\begin{eqnarray}
&&G_D^{-+}(\omega) = 2i {\Gamma_L \Theta (\mu_L - \omega) + \Gamma_R \Theta (\mu_R - \omega) \over
(\omega-\epsilon_0)^2 + (\Gamma_L+\Gamma_R)^2}\,,\\
\label{a4a} &&G_D^{+-}(\omega) = -2i {\Gamma_L \Theta (\omega-\mu_L) + \Gamma_R \Theta
(\omega-\mu_R) \over (\omega-\epsilon_0)^2 + (\Gamma_L + \Gamma_R)^2}\, , \label{a4b}
\end{eqnarray}
\end{mathletters}
where $G_D^{ij}(\omega) = \int G_D^{ij}(t)\exp{(i\omega t)}dt$. In Eq.~(\ref{a4}) we introduced
tunnelling rates $\Gamma_{L(R)} = \pi T^2_{L(R)}\rho_{L(R)}$, where the densities of states in the
leads $\rho_{L(R)}$ are assumed constant for simplicity.

The $O(\lambda^2)$ contribution to the effective action in Eq.~(\ref{a3}) is
generated by the integral kernel $K(t_1-t_2)$, which is just a two-point
correlation function of charge fluctuations in the dot, and can be expressed as
a product of two single particle Green's functions $G_D$. The double integral
in Eq.~(\ref{a3}) can be rewritten as
\begin{eqnarray}
\int_{-\infty}^{\infty}dt_1dt_2\Big\{2i x^c(t_1)x^q(t_2)\Theta(t_1-t_2) A(t_1-t_2)\nonumber\\+
x^q(t_1)x^q(t_2)S(t_1-t_2)\Big\}\, ,\label{a5}
\end{eqnarray}
\noindent where we have introduced the ``rotated'' Keldysh variables $x^q(t) = x(t^+)-x(t^-)$,
$x^c(t) = x(t^+)+x(t^-)$. The kernels $A(t_1-t_2) = {\rm Im} [G_D^{+-} (t_2-t_1)
G_D^{-+}(t_1-t_2)]$ and $S(t_1-t_2) = {\rm Re} [G_D^{+-}(t_2-t_1) G_D^{-+} (t_1-t_2)]$, with
$G_D^{+-}$ and $G_D^{-+}$ given by Eqs.~(\ref{a3}), are related to anti-symmetric (quantum)
and symmetric (classical) parts the charge correlation function $K(t_2-t_1)$.

If we assume that the tunneling through the resonant level is fast compared to
the motion of the oscillator, i.e., the kernel $K(t_2-t_1)$ is nonzero on a
time scale which is much smaller than the time scale of the oscillator, then
the oscillator trajectory between $t_1$ and $t_2$ is close to a straight line,
i. e.,
$x^{c(q)}(t_2) \simeq x^{c(q)}(t_1) + {\dot x}^{c(q)}(t_1)(t_1-t_2)$.
Using this approximation in Eq.~(\ref{a5})
gives an action which is local in time:
\begin{eqnarray}
{\cal F}=-{\lambda^2 \over 2}\int dt \left(i R x^c x^q + i A {\dot x}^c x^q + S x^q x^q\right) +...
\, ,\label{a7}
\end{eqnarray}
where the the coefficients $R$, $A$ and $S$ can be expressed in terms of single particle Green's
functions as follows:
\begin{mathletters}
\label{a8}
\begin{eqnarray}
&&S={1\over 2\pi} \int d\omega \, G_D^{-+}(\omega)G_D^{+-}(\omega)\,,\\
\label{a8a} && A= {1\over 2\pi} \int d\omega \, G_D^{-+}(\omega) {\partial \over \partial\omega}
G_D^{+-}(\omega)\,,\\
\label{a8b} && R = {1\over \pi} \, {\cal P}\int d\omega_1d\omega_2
\,{G_D^{-+}(\omega_1)G_D^{+-}(\omega_2)\over \omega_1-\omega_2}\, .\label{a8c}
\end{eqnarray}
\end{mathletters}

The effective action in Eq.~(\ref{a7}) is exactly of the form of the Caldeira
and Leggett action ~\cite{Caldeira}, derived for a bosonic heat bath at high
temperature. The first term in the effective action ${\cal F}$ is a
renormalization of the oscillator potential. In contrast to the infinite
renormalization in the Caldeira-Leggett model, here
$R$ is finite. Evaluating
the integral in Eq.~(\ref{a8c}) using
Eqs.~(\ref{a4}) yields $R =
2\partial\langle n_0 \rangle_{\rm el}/\partial\epsilon_0$, where $\langle n_0 \rangle_{\rm el}$ is
the average occupation of the electron in the dot given by Eq.~(\ref{a5}). Physically this
corresponds to the back action of tunnel current, which is perturbed by the displacement of the
oscillator.
Combining
this second order renormalization with the first order renormalization in Eq.~(\ref{a3}) the
effective potential of the oscillator can be written as $V(x) \simeq V^{\rm bare}(x) + \lambda x
\langle n_0(x) \rangle_{\rm el}$, where $\langle n_0(x) \rangle_{\rm el}$ is the occupation number
of the resonant level for a fixed position of the oscillator. The last term in
the exponent of Eq.~(\ref{a7}) provides dephasing of the system. Its effect at
the classical level corresponds to a white noise force $f(t)$ exerted by the
tunnel current on the oscillator. The second term causes energy damping. The
classical equation of motion for the oscillator\cite{Caldeira} can be written
as $m{\ddot x} + m\gamma {\dot x} +\partial_x V = f(t)$, where, in our case,
$m\gamma = \lambda^2 A$ and $\langle f(t)f(t^\prime)\rangle = \lambda^2
S\delta(t-t')$. Thus the classical and quantum parts of the resonant level
charge correlation function determine fluctuations and dissipation for the
oscillator respectively. One can therefore define an effective temperature,
$T_{\rm eff}$, using a fluctuation-dissipation relation, giving $S/A = 2T_{\rm
eff}$. The effective temperature, $T_{\rm eff}$, is not determined by the
reservoir's actual temperature, as in the Caldeira-Leggett
model~\cite{Caldeira}, but rather by the coupling to the tunnel current.

$T_{\rm eff}$ determines the
fluctuations of the oscillator coordinate due to the tunnel
current induced noise. In the case of a linear oscillator, $V(x) = m\omega_{\rm
osc}^2 x^2/2$, the dispersion of the oscillator coordinate is $\langle x^2
\rangle= T_{\rm eff}/(m\omega_{\rm osc}^2)$.
We can now check the validity of our expansion in $\lambda$.  From the
structure of Eqs.~(\ref{a7},\ref{a8}), we see that higher order terms in
Eq.~(\ref{a3}) will be smaller by powers of the dimensionless parameter
$\lambda \sqrt{\langle x^2\rangle} /(\Gamma_L+\Gamma_R)$. Physically, if the
oscillator induced shift of the resonant level is small compared to the width
of the level, the back action on the oscillator is weakly dependent on the
position of the oscillator and the higher order nonlinearities are unimportant.
Thus, for sufficiently large $\Gamma$, we only need to consider the leading,
quadratic, terms in the effective action (\ref{a7}).

It is instructive to evaluate $T_{\rm eff}$ explicitly, using
Eqs.~(\ref{a4},\ref{a8}) in two limiting cases--the {\em threshold} and {\em
co-tunneling} regimes.  Suppose first that the resonant level is in the
vicinity of one of the chemical potentials in the leads, say
$\epsilon_0 = \mu_R = 0$ and the bias between the chemical potentials is large,
$\mu_L = \infty$. In this regime, the current through the device is very sensitive to the
energy of the resonant level, and hence to the displacement of the cantilever.
In this case we obtain:
$\gamma^{\rm thr} = \hbar\lambda^2 \Gamma_R /(\pi m\Gamma^3)$,  $T_{\rm eff}^{\rm thr}
= {\pi\Gamma_L/4}$.  Therefore, the effective temperature of the oscillator in
the threshold regime is essentially defined by the tunneling-induced width
$\Gamma$ of the resonant level. For a practical Si nano-mechanical resonator
with dimensions 3$\mu$m $\times$ 0.1$\mu$m $\times$ 0.1$\mu$m, coupled to SET
with $\Gamma\sim 10^9$ s$^{-1}$ by a coupling strength
$\lambda\sim 10^{-3}$V/\AA, this corresponds to an effective temperature
$T_{\rm eff}\sim 0.1$ K, and the damping coefficient $\gamma\sim 10^{7}$
s$^{-1}$.  The effect of back action in this case will limit the lowest
achievable oscillator temperature to 0.1 K, and the maximum quality factor to
about 100.

Another important limiting regime corresponds to the situation when the
resonant level is far above or below the chemical potentials in the leads. This
is the so-called co-tunneling regime. The tunneling electrons can now occupy
the level only virtually and the effective coupling constant between the leads
is small as it is suppressed by the large energy separation between the
chemical potentials in the leads and the resonant level.  Assuming
$\mu_L-\mu_R \equiv V \ll (\mu_L+\mu_R)/2
\equiv \mu_F$, and $\mu_L(R)\gg \Gamma, \epsilon_0$, we obtain
$\gamma^{\rm cot} = \hbar\lambda^2 \Gamma^2 / (\pi m\mu_F^4)$, $T_{\rm eff}^{\rm cot}
= {\Gamma_L\Gamma_R V / \Gamma^2}$.  Compared to the threshold regime, in the
co-tunneling both the oscillator damping and the back-action noise are
significantly reduced, with the bias voltage across the leads determining the
effective temperature of the oscillator. This result is consistent with
Ref.~\cite{MM,HMM}.

We are now in a position to analyze the sensitivity of the system. Suppose that
the oscillator (which will be assumed linear from now on) is perturbed by an
external force $F(t)$, say, a short kick of duration
$\tau_F\ll\omega_0^{-1}$, so that $F(t) \simeq F\tau_F \delta (t)$.  This kick
results in the variation of the oscillator's amplitude by the amount $\delta x
= F\tau_F/(m\omega_0)$. What minimum $\delta x$ can be detected by observing
the tunnel current, given the noise $\langle |I_\omega|^2\rangle$ in the
current?

\begin{figure}
{\centering{\psfig{figure=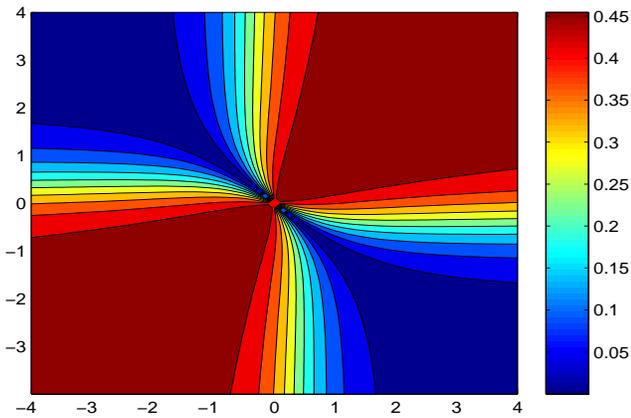,height=2.2 in,width=3.3
in}}}\caption{Sensitivity in dimensionless units ($x_0^2/\delta x_{\rm min}^2$)
as a function of chemical potentials for a symmetric structure
($\Gamma_L=\Gamma_R=\Gamma/2$). The axes are $\mu_L/\Gamma$ and
$\mu_R/\Gamma$.}
\end{figure}

The ability to measure a signal can be represented by the integrated signal-to-noise
ratio~\cite{Braginsky}
\begin{equation}
s/n=(1/2\pi)\int d\omega |\langle S_F(\omega) \rangle|^2/\langle |I_\omega|^2\rangle\, ,\label{a11}
\end{equation}
where $S_F(\omega)$ is the Fourier transform of the detector's response to an
external perturbation, i.e., the force $F$. The variation of the current
through the structure due to the variation of the oscillator coordinate is
$\delta I = (\partial I /\partial x)\delta x$, and therefore the response can
be written as $S_F(\omega)=(\partial I/\partial
x)F(\omega)/[m(\omega^2+i\gamma\omega-\omega_0^2)]$.

The current and the noise can be easily evaluated if we recall that the
dynamics of the oscillator is slow compared to the dynamics of the tunneling
electrons. The current in the adiabatic approximation, i.e., for a fixed
position of the oscillator is given by $I(x) =(1/2\pi)\int_{\mu_R}^{\mu_L}
{d\omega T(\omega, x)}$, where the transmission coefficient $ T(\omega, x)=
4\Gamma_L\Gamma_R/[(\omega - E_0(x))^2+\Gamma^2]$~\cite{Juaho}. The noise at
low frequencies (of order $\omega_0$) is given by $\langle |I_\omega|^2\rangle
\simeq \langle |I_0|^2\rangle + (\partial I/\partial x)^2\langle |\Delta
x_\omega|^2\rangle$, where the shot noise is related to the transmission
coefficient as $\langle |I_0|^2\rangle = (1/2\pi)\int_{\mu_R}^{\mu_L} {d\omega
T(\omega, 0)[1-T(\omega, 0)]}$~\cite{Lesovik}, and $\langle |\Delta
x_\omega|^2\rangle = 2\gamma T_{\rm eff}/m[(\omega^2-\omega_0^2)^2 +
\gamma^2\omega^2]$ is the fluctuation spectrum for the oscillator. Substituting
these expressions into Eq.~(\ref{a11}) and setting the $s/n=1$ as a criterion
for a successful measurement, one obtains a criterion for a detection of a
minimum force by our apparatus. By expressing this minimum force in terms of
the minimum displacement $\delta x_{\rm min}$ that it causes, this criterion
reduces to
\begin{equation}
\label{sens}
x_0^2/\delta x_{\rm min}^2 \simeq |\partial I/\partial \epsilon_0|/\left(4\sqrt{S\langle
|I_0|^2\rangle}\right)\, ,\label{a12}
\end{equation}
where $x_0^2 = \hbar/(2m\omega_0)$ is zero point displacement for the oscillator, and $S$ is given
by Eq.~(\ref{a8a}).

The sensitivity defined by Eq.~(\ref{sens}) can be evaluated and is presented
in Fig. 2 as a function of chemical potentials in the leads relative to the
position of the resonant level. The sensitivity is maximal in the co-tunneling
regime, where it reaches $1/2$. In the threshold regime the sensitivity is
somewhat smaller ($\simeq$ by a factor of 2). The sensitivity is worst when the
resonant level is positioned symmetrically with respect to the chemical
potentials (the blue regions in Fig. 2), as the current sensitivity $\partial
I/\partial x$ vanishes in this regime.  These results are similar to the
conclusions reached in the studies of the ``non-ideality'' of SET
detectors\cite{Averin}. Qualitatively, the reduction of sensitivity in the
sequential tunneling regime can be attributed to the detector latency during
the electron dwell time on the island, which contributes to the back-action
noise, but not to the measurement.

In summary, we have analyzed the quantum measurement of a mechanical oscillator
coupled to an electronic resonant level that models a single electron
transistor.  We determined the back action effects of the detector on the
quantum system, which lead to a measurement-induced effective temperature and
damping coefficient.  We also determined the fundamental sensitivity limits of
the scheme in all operation regimes.

We would like to acknowledge useful discussions with K. Schwab and D. Pelekhov.
This work was supported by the US DOE. D.M. was supported, in part, by the US
NSF grants ECS-0102500 and DMR-0121146.  MBH was supported by the DOE grant
W-740S-ENG-36.

\end{multicols}
\end{document}